\documentclass[twocolumn,amsmath,amssymb]{revtex4}
\usepackage{graphicx}
\usepackage{color}

\begin{document}

\title{Phase diagram and universality of the Lennard-Jones gas-liquid system}

\author{Hiroshi Watanabe$^{1}\footnote{E-mail: hwatanabe@issp.u-tokyo.ac.jp.}$,
Nobuyasu Ito$^{2}$,
and Chin-Kun Hu$^{3}$\footnote{E-mail: huck@phys.sinica.edu.tw.}}

\affiliation{$^1$
The Institute for Solid State Physics, The University of Tokyo,
Kashiwanoha 5-1-5, Kashiwa, Chiba 277-8581, JAPAN
}

\affiliation{
$^2$ Department of Applied Physics, School of Engineering,
The University of Tokyo, Hongo, Bunkyo-ku, Tokyo 113-8656, Japan
}

\affiliation{$^3$Institute of Physics, Academia
Sinica, Nankang, Taipei 11529, Taiwan}

\date{\today}

\begin{abstract}
The gas-liquid phase transition of the
three-dimensional Lennard-Jones particles system is studied by
molecular dynamics simulations. The gas and liquid
densities in the coexisting state are determined
with high accuracy. The critical point is determined by the block
density analysis of the Binder parameter with the aid of the law
of rectilinear diameter.
From the critical behavior of the gas-liquid coexsisting density,
the critical exponent of the order parameter is estimated to be
$\beta = 0.3285(7)$. Surface tension is estimated from interface
broadening behavior due to capillary waves. From the critical
behavior of the surface tension, the critical exponent of the
correlation length is estimated to be $\nu = 0.63 (4)$.
The obtained values of $\beta$ and $\nu$ are
consistent with those of the Ising universality class.
\end{abstract}


\maketitle

\section{Introduction}

University and scaling are key concepts in the
study of critical systems \cite{71Gene,90Leo}, including
liquid-gas systems \cite{45jcp,beta,09jcpBinder}, Ising model
\cite{44ising,97ising}, percolation model
\cite{sa94,95prl,96prl,04prl,05prlwh}, dimer model \cite{dimer},
etc. According to modern theory of critical phenomena
\cite{71Gene,90Leo}, critical systems can be classified into
different universality classes such that the systems in the same
class have the same set of critical exponents.

The Lennard-Jones (LJ) particle system has been extensively
studied since it is the basic model of the off-lattice systems
with short-range interactions. It is widely believed that the
gas-liquid phase transition of the particle system with
short-range interaction belongs to Ising universality class
\cite{52pr-LeeYang} and the LJ-particle system is no exception.
However, there is no definite proof that the universality class of
the liquid-gas phase transition of the LJ system is same as that
of the Ising universality. In order to study the universality
class of the LJ system, the precise information of the phase
diagram, especially the location of the critical point, is
required. Aside from the critical behavior, the determination of
the phase boundary itself is important to study nonequilibrium
phenomena such as bubble nuclei, etc~\cite{Watanabe2010}.

Recently, due to the advance in the computational
power, many researchers have tried to determine the value of the
critical exponents as accurate as possible, in order to confirm
whether the universality class of the LJ system is really the same
as that of the Ising model. Panagiotopoulos proposed the Gibbs
ensemble Monte Carlo (GEMC) method to calculate the phase
equilibria of the coexisting phase~\cite{Panagiotopoulos}. The
GEMC method involves two simulation boxes, one is in liquid phase
and the other is in gas phase under a given temperature. There
are three types of Monte Carlo (MC) steps; displacements of
particles, changing volumes of simulation boxes, and exchanging
particles between the simulation boxes. When the two systems reach
their equilibrium states, we can obtain the physical quantities
such as densities of gas and liquid phases, vapor pressure and
chemical potential of the coexisting state in the given
temperature. While this method is simple, the system can be
unstable near the criticality, \textit{e.g.}, the simulation box
may change its phase from gas to liquid and vice versa because of
the fluctuation.

M\"oller and Fischer proposed a different approach, $NpT$ and test
particle method~\cite{Moeller}. This method involves $NpT$
simulation, \textit{i.e.}, the isobaric and isothermal condition.
The chemical potential is observed as a function of temperature by
Widom's test-particle method~\cite{Widom}, and the critical point
is determined by the crossing point of the chemical potential of
liquid and gas branches. This method is more stable than GEMC near
the critical point; since only the volume
fluctuates while both density and volume fluctuate in GEMC.
Additionally, Okumura and Yonezawa \cite{Okumura} proposed $NVT$
and test particle method which is more stable than the last two
methods since this method does not involve change in volume.
However, the calculation of the chemical potential with Widom's
test-particle method is quite expensive, especially for larger
systems. Wilding and Bruce combined the histogram reweighting (HR)
techniques with the finite-size scaling (FSS)
analysis~\cite{Wilding}. The main idea of this method is to locate
the apparent critical point from the matching condition between
the ordering operator distribution and that of Ising model with
the help of the mixed-field theory.
P\'erez-Pellitero \textit{et al.}\ compared the HR techniques with the direct estimation
of the critical point using Binder parameter~\cite{Pellitero}.

While the HR techniques combined with FSS analysis are efficient to study the coexisting phase,
the distribution of the order parameter of Ising model at the critical point is required for the reference.
Additionally, most of past studies determined the critical point with the help of the
knowledge of the Ising universality, and therefore, it is difficult to determine
whether the gas-liquid phase transition of LJ-particle system belongs to the Ising universality class.

In the present study, we propose a simple method to study the
gas-liquid coexisting phase using the molecular dynamics (MD)
simulations. The main purpose of our method is to determine the
critical properties of the LJ gas-liquid system without \textit{a
priori} knowledge of the Ising universality, such as the values of
the critical exponents, the value of Binder parameter at the
critical point, or the distribution of the order parameter, etc.
The critical exponents of the order parameter is estimated from
the liquid-gas coexisting  density and the that of the correlation
length is estimated from the interface thickness taking into
account the effect of the capillary waves. We have
found that the critical exponents of the order parameter and
correlation length are $\beta = 0.3285(7)$ and $\nu = 0.63(4)$,
respectively, which are consistent with those of the Ising
universality class~\cite{71Gene,96jpaIsing,Ito2000,Hasenbusch}.

This paper is organized as follows. In Sec.~II, we
describe our model system and details of the simulation procedure.
We first describe how to determine the coexisting curve in Sec.~II
A, then we describe how to locate the critical point using the
Binder parameter with the aid of the law of rectilinear diameter
in Sec.~II B. In Sec.~III, we present results of our simulation
data and critical exponents determined from such data. Sec.~IV
contains our conclusions and discusses problems for further
studies.

\begin{figure}[htbp]
\begin{center}
\includegraphics[width=8.5cm]{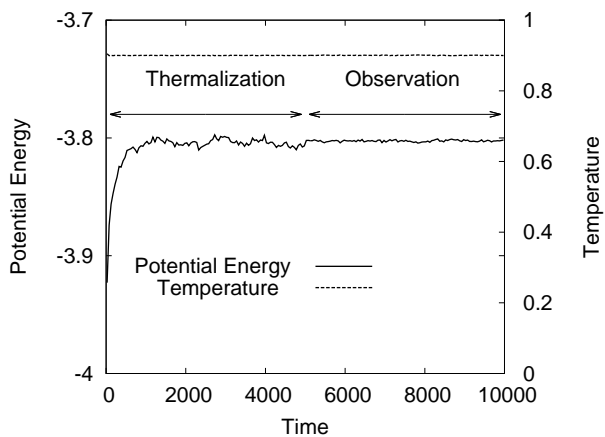}
\end{center}
\caption{ Relaxation behaviors of potential energy (solid line)
and temperature (dashed line). The system size is $64 \times 64
\times 128$ with the number of particles 180288. Aimed temperature
is 0.9. While the relaxation of temperature is quick, that of the
potential energy is slow. After thermalization, the heatbath is
turned off. } \label{fig_relaxation}
\end{figure}

\section{Model system and simulation procedure}

\subsection{Model system and coexisting curve}

We use the truncated Lennard-Jones potential of the form
\begin{equation}
V(r) =
\displaystyle 4 \varepsilon \left[
\left( \frac{\sigma}{r} \right)^{12} -
\left( \frac{\sigma}{r} \right)^{6} +
c_2 \left( \frac{r}{\sigma} \right)^2+ c_0 \right],
\end{equation}
with the well depth $\varepsilon$ and atomic diameter
$\sigma$~\cite{Spotswood1973}. The coefficients $c_2$ and $c_0$
are determined so that $V(r_c) = V'(r_c) = 0$ with the cut-off
length $r_c$ , i.e., the values of potential and force become
continuously zero at the truncation point and
$V(r)=0$ for $r > r_c.$
The two additional terms are introduced for good
conservation of the total energy. The extra terms, which
makes the force continuous at the truncation point,
are required to perform the MD simulations stably for long time
and it is unnecessary for MC simulations.
In the following, we use the physical quantities reduced by
$\sigma$, $\varepsilon$, and the Boltzmann constant
$k_\mathrm{B}$, e.g., the length is measured with
the unit of $\sigma$, and so forth. We set the cutoff length as
$r_c = 3.0$. The simulation box is a rectangular parallelepiped
with sizes $L_x \times L_y \times L_z$. The
periodic boundary condition is taken for all directions. We set
the ratio of $L_x:L_y:L_z = 1:1:2$ in order to achieve the
gas-liquid coexisting phase. The studied system sizes are $L_x =
32, 64, 96,$ and  $128$. We setup the initial configuration by
placing particles densely in the region $z <L_z/2$ and sparsely in
the region $z \geq L_z/2$, and we equibrate the system with the
heatbath. Nos\'e--Hoover heatbath is used for controlling
temperature~\cite{NoseHoover}. Time integration scheme is the
second order RESPA~\cite{Tuckerman1992} for isothermal simulations
and the second order symplectic integrator for isoenergetic
simulations. The time step is chosen to be $0.005$ throughout the
simulations. We check whether the system has reached the
equilibrium state by observing the potential energy, since the
relaxation of the potential energy is much slower than that of the
temperature (see Fig.~\ref{fig_relaxation}). The potential energy
$U(t)$ behaves as $U(t) -  U_\mathrm{eq}  \propto \exp(-t/\tau)$
with the equilibrium value $U_\mathrm{eq}$ and the relaxation time
$\tau$. We perform thermalization at least ten times longer than
the relaxation time. Then we turn off the heatbath and start
observation of the density profile at and near
gas-liquid phase interface and surface tension. The simulations
are performed with MPI parallelized MD program~\cite{mdacp,
mdnote}. The largest run, involving 1583504 particles and 5000000
steps, takes about 8 hours using 1024 cores at SGI Altix ICE 8400 EX.

\begin{figure}[htbp]
\begin{center}
\includegraphics[width=8.5cm]{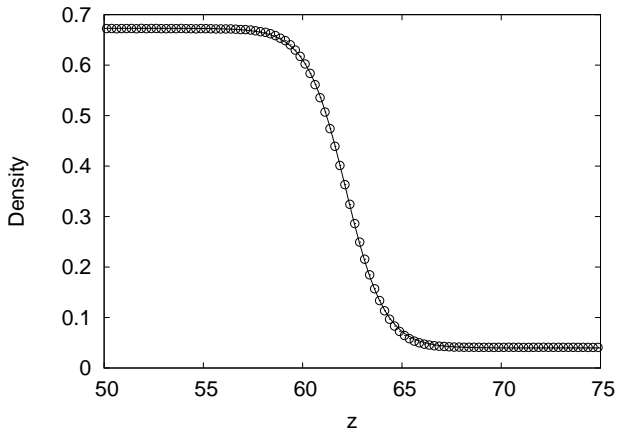}
\end{center}
\caption{Density profile at and near the gas-liquid interface. The
conditions of the simulation are same as
Fig.~\ref{fig_relaxation}. The open circles are the simulation
results and the solid line denotes the fitting result via
Eq.~(\ref{eq_tanh}) with parameters $\rho_\mathrm{l}=0.6729(3)$,
$\rho_\mathrm{g}= 0.0397(3)$, $z_c=62.161(4)$, and $\lambda=
1.956(7) $. Statistical errors are smaller than the symbol size. }
\label{fig_density}
\end{figure}

\begin{figure}[tbp]
\begin{center}
\includegraphics[width=8.4cm]{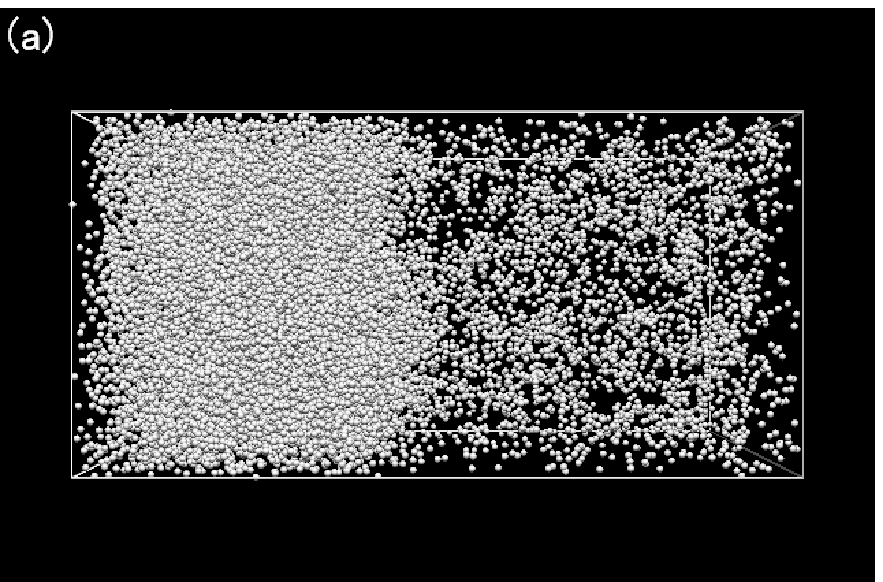}
\includegraphics[width=8.4cm]{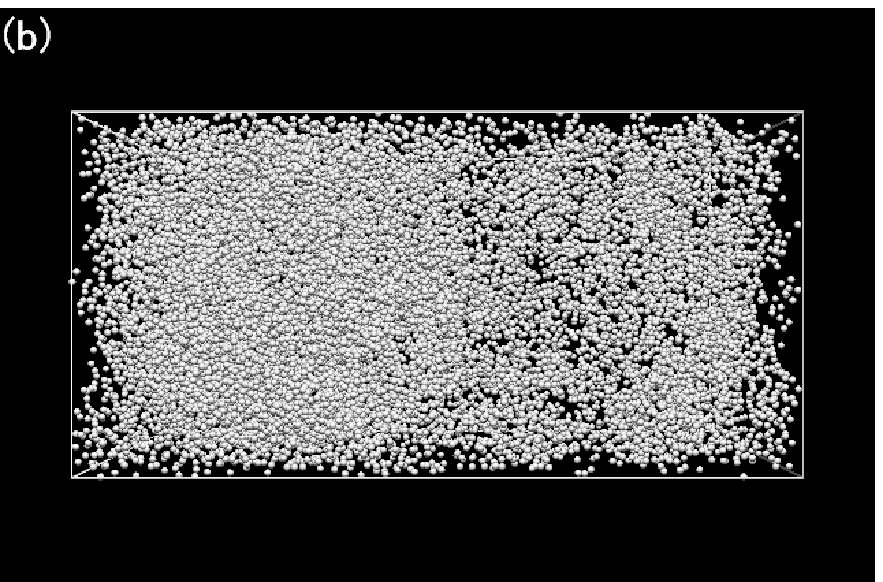}
\includegraphics[width=8.4cm]{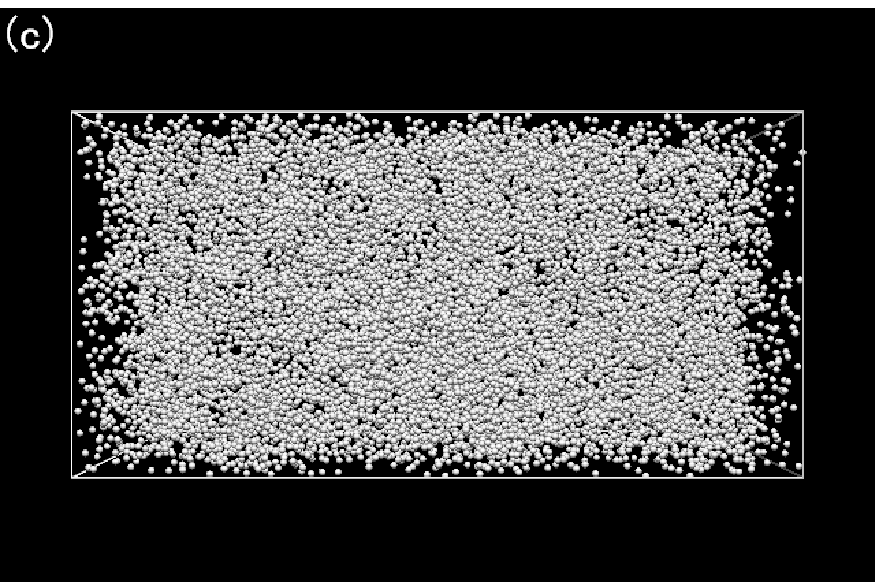}
\end{center}
\caption{
Typical configurations of particles.
Small systems with sizes $(L_x, L_y, L_z) =  (32, 32, 64)$ are shown for the visibility.
(a) Below critical points ($T=1.00$) with 18432 particles. The gas (left) and liquid (right) phases are clearly separated.
(b) Near critical points ($T=1.09$) with 18824 particles. The correlation length becomes longer and the phase interface becomes unclear.
(c) Above critical points ($T=1.11$) with 17576 particles. The system becomes uniform.
}
\label{fig_snapshots}
\end{figure}

The functional  form of the density profile at and
near  the gas-liquid interface is of the form,
\begin{equation}
\rho(z) = \frac{(\rho_\mathrm{l} - \rho_\mathrm{g})}{2}
\left(\tanh{\left[(z_c-z)/\lambda\right]}+1\right) +
\rho_\mathrm{g}, \label{eq_tanh}
\end{equation}
where $\rho_\mathrm{l}$, $\rho_\mathrm{g}$, and $\lambda$ are
the density of liquid, the density of gas, and the thickness of the interface.
The typical density profile is shown in Fig.~\ref{fig_density}.
By fitting Eq.~(\ref{eq_tanh}) to the simulation results, we obtain the
gas density $\rho_\mathrm{g}$, the liquid density $\rho_\mathrm{l}$, and the
thickness $\lambda$ as functions of temperature.
The  thickness $\lambda$ is associated with the correlation length, and becomes
longer as the temperature approaches to the critical point.
The finite-size effect can be ignored while $\lambda$ is sufficiently shorter
than the system size $L_x$.
This means that larger system size allows us to approach closer to the critical point.
Typical configurations of particles below, near,
and above the critical point are shown in Figs.~\ref{fig_snapshots}(a), \ref{fig_snapshots}(b), and
\ref{fig_snapshots}(c), respectively.

\subsection{Surface tension}
When the phase interfaces are normal to the $z$-axis,
the surface tension $\gamma$ is defined by the following equation
\begin{equation}
2 \gamma L_x L_y = \int \mathrm{d} \textbf{r} \left(
\pi_{zz}(\textbf{r}) - \frac{\pi_{xx}(\textbf{r})  + \pi_{yy}(\textbf{r}) }{2}
\right), \label{eq_stress}
\end{equation}
where $\pi_{xx}$, $\pi_{yy}$ and $\pi_{zz}$ are the diagonal components of the stress tensor~\cite{McLennan}.
The integral taken over the whole volume of the simulation box.
The factor $2$ of the l.h.s.\ comes from the fact that there are two interfaces in the system
due to the periodic condition.
For the particles interacting via two body potential $V(r)$, Eq.~(\ref{eq_stress})
reduces as
\begin{equation}
2 \gamma L_x L_y = \sum_{i < j} \frac{x_{ij}^2 + y_{ij}^2 - 2 z_{ij}^2 }{2} V'(r_{ij}), \label{eq_gamma}
\end{equation}
where $r_{ij}$ is the distance between particles
$i$ and $j$, $V'(r_{ij})$ is the derivative of $V(r_{ij})$ with
respect to  $r_{ij}$, $x_{ij}$ is the $x$ component of the
relative position vector between particles $i$ and $j$, and so
forth. The sum is taken over all pairs. Note that, the kinetic
term of the stress tensor is omitted since the kinetic term does
not contribute to the surface tension. After the thermalization,
the surface tension is sampled every 1000 steps and averaged. The
observation is performed for 100000 steps for each temperature.

\subsection{Critical behavior}

The order parameter of the gas-liquid transition is the density difference
$\Delta \rho \equiv \rho_\mathrm{l} - \rho_\mathrm{g}$.
The order parameter becomes zero at the critical temperature $T_c$, and
the behavior near the criticality is
\begin{equation}
\Delta \rho \sim \varepsilon^\beta,
\end{equation}
with the reduced temperature $\varepsilon \equiv  |T_c-T|/T_c$ and the
critical exponent $\beta$.
The surface tension $\gamma$ and the thickness of the interface $\lambda$
also show critical behavior as follows,
\begin{eqnarray}
\gamma &\sim& \varepsilon^{2\nu}, \\
\lambda &\sim & \varepsilon^{-\nu},
\end{eqnarray}
with the critical exponent of the correlation length $\nu$.
It is difficult to determine the critical temperature and the critical exponents simultaneously
only from the above scaling relations without assuming the Ising universality.
Therefore, we determine the critical point using the Binder parameter~\cite{Binder}.

In order to compute the fluctuation of density from $NVT$ or $NVE$ simulations,
we adopt the block distribution analysis~\cite{Rovere}. The main
idea of this method is to divide the system into small cells, and
each cell exhibits approximately the grandcanonical ensemble,
while cells are not completely independent from
each other. Suppose there are a cubic system with linear dimension
$L$ containing $N$ particles. The simulation box with linear
dimension $L$ is divided in to cells of size $L/M_b$, where $M_b$
is an integer. Each cell contains different
number of particles, and therefore, has different density
$\rho_i$, $i$ being the label of the cells. The moments of density required for calculation of Binder parameter
are defined to be,
\begin{eqnarray}
\bar{\rho} &=&  \frac{1}{M_b^3} \sum_i \rho_i, \\
m^2 &=&  \frac{1}{M_b^3} \sum_i  (\rho_i - \bar{\rho})^2, \\
m^4 &=& \frac{1}{M_b^3}  \sum_i  (\rho_i - \bar{\rho})^4.
\end{eqnarray}
Using the above definition, we can define the Binder parameters as follows:
\begin{equation}
U(M_b) =\frac{\left< m^4 \right>}{\left< m^2 \right>^2},
\end{equation}
where $\left< \cdots \right> $ denotes ensemble average. With
different dividing number $M_b$, we can obtain the Binder
parameters for different volume sizes from a
single run for a sufficiently large system.

The crossing point of
the Binder parameter gives the critical point. Since we perform
NVT ensemble, the density and the temperature should be given
simultaneously. Therefore, we assume the law of rectilinear
diameters~\cite{rectilinear} that the average density shows linear
dependence on temperature as  $(\rho_\mathrm{g} +
\rho_\mathrm{l})/2 = a T + b$. While some fluids exhibit strong
deviations from the above linear relation, simple pure fluids such
as oxygen and xeon follow this law within high
accuracy~\cite{Wang2007}. The studied system is a cube with linear
dimension $L=128$. First the system is thermalized for $200000$
steps with heatbath, and the Binder parameters are observed every
1000 steps. The observation is performed for $100000$ steps. The
Binder parameters are calculated for $M_b = 16, 24,$ and $32$.

While the critical exponent can be also estimated
from the finite-size analysis of Binder parameter,
it is better to estimate the critical exponent $\beta$
from the liquid-gas coexisting density.
Because the coexisting density can be determined more accurate
than Binder parameter, since the coexisting density is the
first order moment of the density while Binder parameter
contains the higher moments.
Therefore, we use Binder parameter only for locating
the critical point and we estimate the critical exponent
from the liquid-gas coexisting density.

We first calculate the phase diagram, and determine the coefficients of the linear relation.
Then we calculate the Binder parameter on the line in order to determine the critical temperature.
Using the obtained critical temperature, we determine the critical exponents.

\begin{figure}[htbp]
\begin{center}
\includegraphics[width=8.5cm]{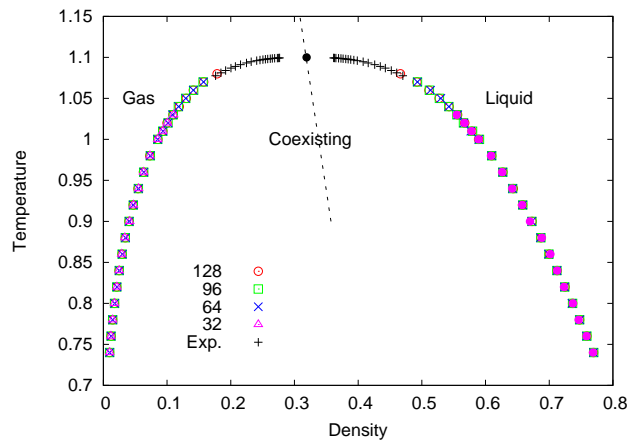}
\end{center}
\caption{ (Color online) Phase boundary between the gas and the
liquid phase obtained by the simulation with $L_x =32, 64, 96$, and $128$.
Statistical errors are smaller than the symbol size. The
experimental result of neon is denoted by
``Exp"~\cite{Pestak1987}. The finite-size effect is hardly
observed. The dashed line denotes the law of rectilinear diameters
and the solid circle denotes the critical point obtained from the
analyses of the Binder parameter. } \label{fig_gl}
\end{figure}

\begin{figure}[htbp]
\begin{center}
\includegraphics[width=8.5cm]{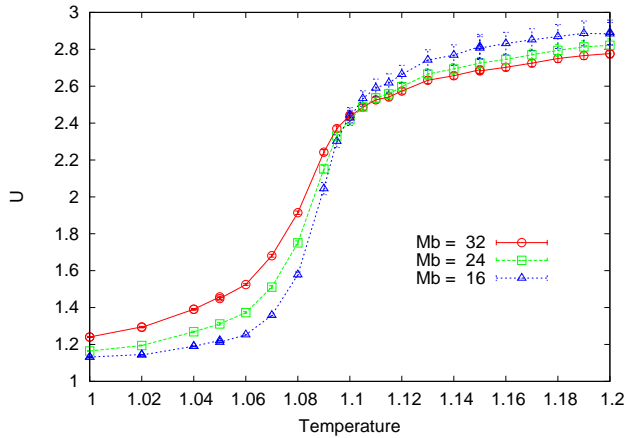}
\end{center}
\caption{ (Color online) Temperature dependence of the Binder
parameter with different block sizes. The system is divided in to
cells of size $L/M_b$, where $L=128$ and $M_b = 16, 24,$ and $32$.
As the value of $M_b$ becomes larger, the size of cells becomes
smaller. From the intersection point, the critical temperature is
estimated to be $T_c = 1.100(5)$. } \label{fig_binder}
\end{figure}

\begin{table}[htb]
\begin{tabular}{|cccccc|}
\hline
$T$& $N$& $\rho_\mathrm{g}$&$\rho_\mathrm{l}$&$\gamma$&$\lambda$\\
\hline
0.74 & 1575720 & 0.0097(2) & 0.7696(3) & 0.55(2) & 1.343(5)\\
0.76 & 1579396 & 0.0120(2) & 0.7591(2) & 0.51(2) & 1.417(4)\\
0.78 & 1583504 & 0.0145(2) & 0.7477(2) & 0.47(2) & 1.457(5)\\
0.8 & 1524992 & 0.0174(2) & 0.7367(2) & 0.436(8) & 1.556(5)\\
0.82 & 1535584 & 0.0209(2) & 0.7249(2) & 0.39(4) & 1.633(5)\\
0.84 & 1541660 & 0.0246(2) & 0.7125(2) & 0.36(1) & 1.734(6)\\
0.86 & 1486940 & 0.0292(2) & 0.7002(2) & 0.32(2) & 1.804(5)\\
0.88 & 1501948 & 0.0344(2) & 0.6870(2) & 0.289(5) & 1.983(6)\\
0.9 & 1450732 & 0.0402(2) & 0.6730(2) & 0.26(2) & 2.110(7)\\
0.92 & 1469556 & 0.0468(2) & 0.6581(2) & 0.22(1) & 2.298(7)\\
0.94 & 1422036 & 0.0543(2) & 0.6436(2) & 0.18(1) & 2.446(7)\\
0.96 & 1445108 & 0.0628(2) & 0.6273(2) & 0.17(2) & 2.630(8)\\
0.98 & 1401476 & 0.0729(2) & 0.6097(2) & 0.14(2) & 2.959(9)\\
1.0 & 1374552 & 0.0856(2) & 0.5891(1) & 0.12(1) & 3.229(8)\\
1.01 & 1389676 & 0.0934(1) & 0.5788(1) & 0.081(7) & 3.483(8)\\
1.02 & 1352596 & 0.1000(2) & 0.5674(1) & 0.08(1) & 3.771(9)\\
1.03 & 1369472 & 0.1092(1) & 0.5562(2) & 0.079(5) & 3.92(1)\\
1.04 & 1335776 & 0.1192(1) & 0.5428(2) & 0.05(1) & 4.46(1)\\
1.05 & 1354500 & 0.1302(1) & 0.5272(1) & 0.034(7) & 4.88(1)\\
1.06 & 1324260 & 0.1412(1) & 0.5112(1) & 0.031(4) & 5.77(1)\\
1.07 & 1318216 & 0.15748(10) & 0.4924(1) & 0.029(10) & 6.97(2)\\
1.08 & 1317812 & 0.17863(9) & 0.46640(8) & 0.016(7) & 8.19(2)\\
\hline
\end{tabular}
\caption{ The data obtained from the simulations with the size
$128 \times 128 \times 256$. The listed quantities are temperature
$T$, the number of particles $N$, the gas density
$\rho_\mathrm{g}$, the liquid density $\rho_\mathrm{l}$, the
surface tension $\gamma$, and the interface thickness $\lambda$,
respectively. } \label{tbl_data}
\end{table}

\begin{figure}[htbp]
\begin{center}
\includegraphics[width=8.5cm]{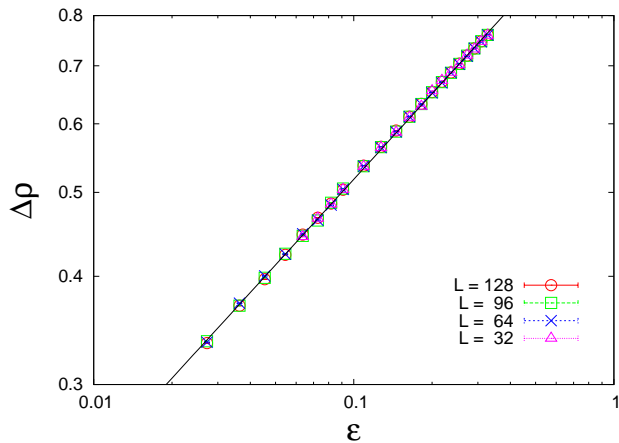}
\end{center}
\caption{ (Color online) Order parameter $\Delta \rho \equiv
\rho_\mathrm{l} - \rho_\mathrm{g}$ is shown as a function of the
reduced temperature $\varepsilon \equiv |T_c - T|/T_c$. The
decimal logarithm is taken for the both axes. Statistical errors
are smaller than the symbol size. Significant differences are
hardly observed between the data of different system sizes. The
critical exponent are determined to be $\beta = 0.3285(7)$.
The solid line is the eye guide with the slope $\beta = 0.3285$.
}
\label{fig_deltarho}
\end{figure}

\begin{figure}[htbp]
\begin{center}
\includegraphics[width=8.5cm]{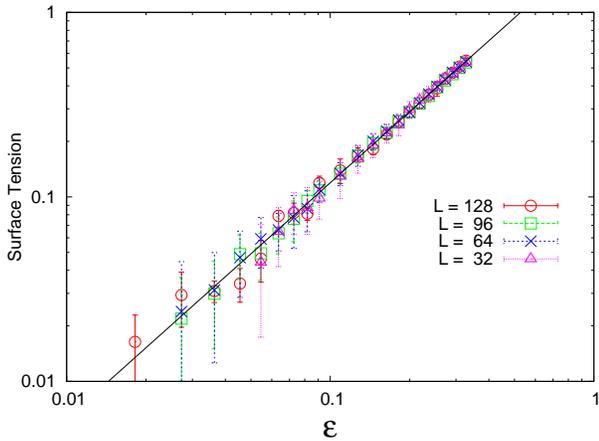}
\end{center}
\caption{(Color online) Temperature dependence of the surface
tension. The decimal logarithm is taken for both
axes. The critical exponent is determined to be $\nu = 0.640(4)$.
The solid line is the eye guide with the slope $2\nu = 1.28$.
} \label{fig_gamma}
\end{figure}

\section{Results}

The gas-liquid phase boundary obtained from the simulations are shown in Fig.~\ref{fig_gl}.
The obtained gas and liquid coexisting densities do not exhibit
finite-size effect, \textit{i.e.}, the values of different system sizes are within the statistical errors.
We also include the experimental data of neon which is denoted by ``Exp"~\cite{Pestak1987}.
The obtained results for the largest systems $L_x =128$ are listed in Table.~\ref{tbl_data}.
We confirm that the law of rectilinear diameters $(\rho_\mathrm{g} + \rho_\mathrm{l})/2 = a T + b$
is well satisfied. We determine the coefficients to be $a = -0.195(1)$ and $b = 0.533(1)$.
The obtained line is shown as the dashed line in Fig.~\ref{fig_gl}.
Note that, we extend the line into the supercritical phase while the
law of rectilinear diameters does not make sense outside the coexisting phase,
since we want to calculate binder parameter
in the region over the critical point in order to have the intersection point.

The Binder parameters are calculated on this line and they are
shown in Fig.~\ref{fig_binder}. The Binder parameter does not
intersect exactly at one point as reported
before~\cite{Pellitero}, but we can determine the critical
temperature with reasonable accuracy as $T_c = 1.100(5)$. Using
the critical point, we determine the value of the critical
exponent $\beta$ from the slope of $\ln \Delta \rho/ \ln
\varepsilon$ as shown in Fig.~\ref{fig_deltarho}. The obtained
value $\beta = 0.3285(7)$ is consistent with the reported value
$\beta=0.3269(6)$ of the spontaneous magnetization
\cite{71Gene} of the three-dimensional Ising model on the simple
cubic lattice obtained by Talapov and Blote \cite{96jpaIsing} or
$\beta =0.325(5)$  by Ito, {\it et al.}~\cite{Ito2000} of the
Ising universality class. The critical behavior of the surface
tension is shown in Fig.~\ref{fig_gamma}. The critical exponent of
the correlation length is determined to be $\nu =0.640(4)$ which
is slightly larger than that of the Ising
universality class $\nu = 0.63002(10)$~\cite{Hasenbusch}.

\begin{figure}[htbp]
\begin{center}
\includegraphics[width=8.5cm]{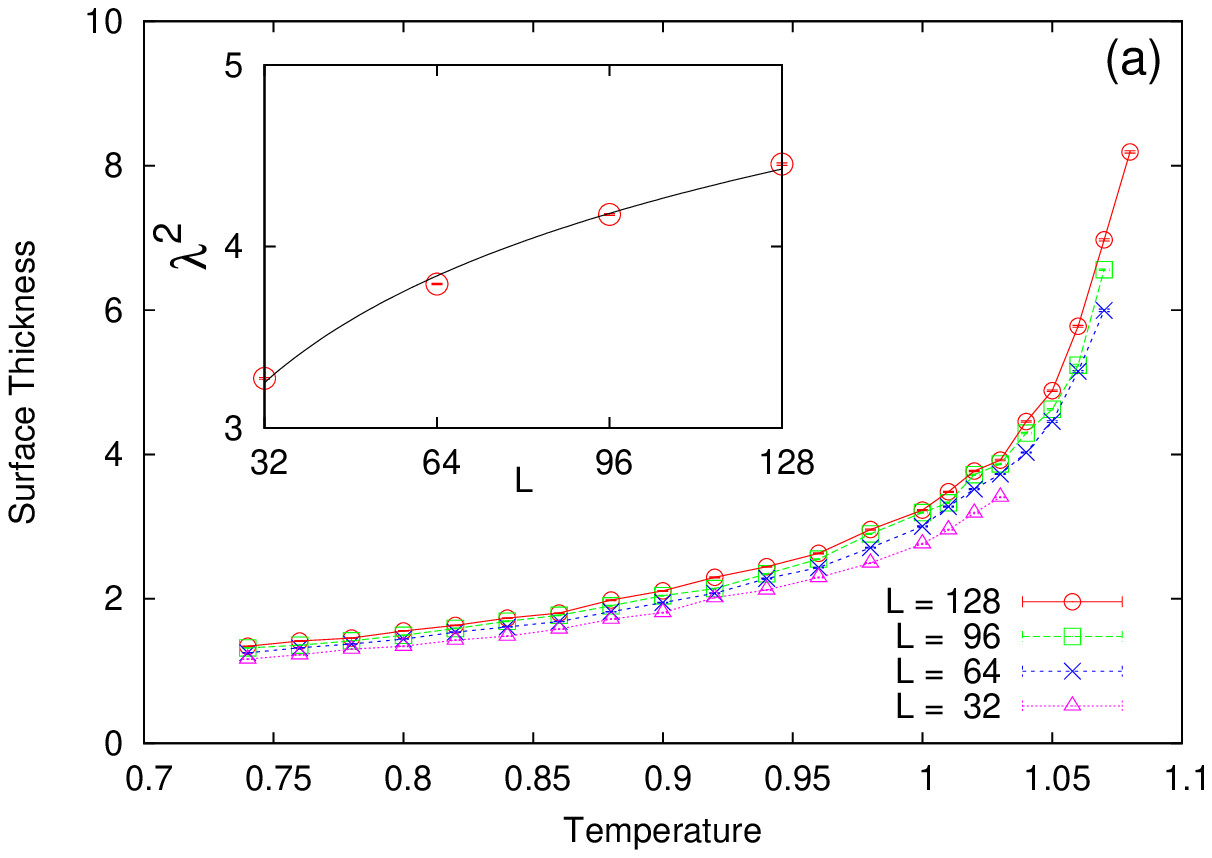}
\includegraphics[width=8.5cm]{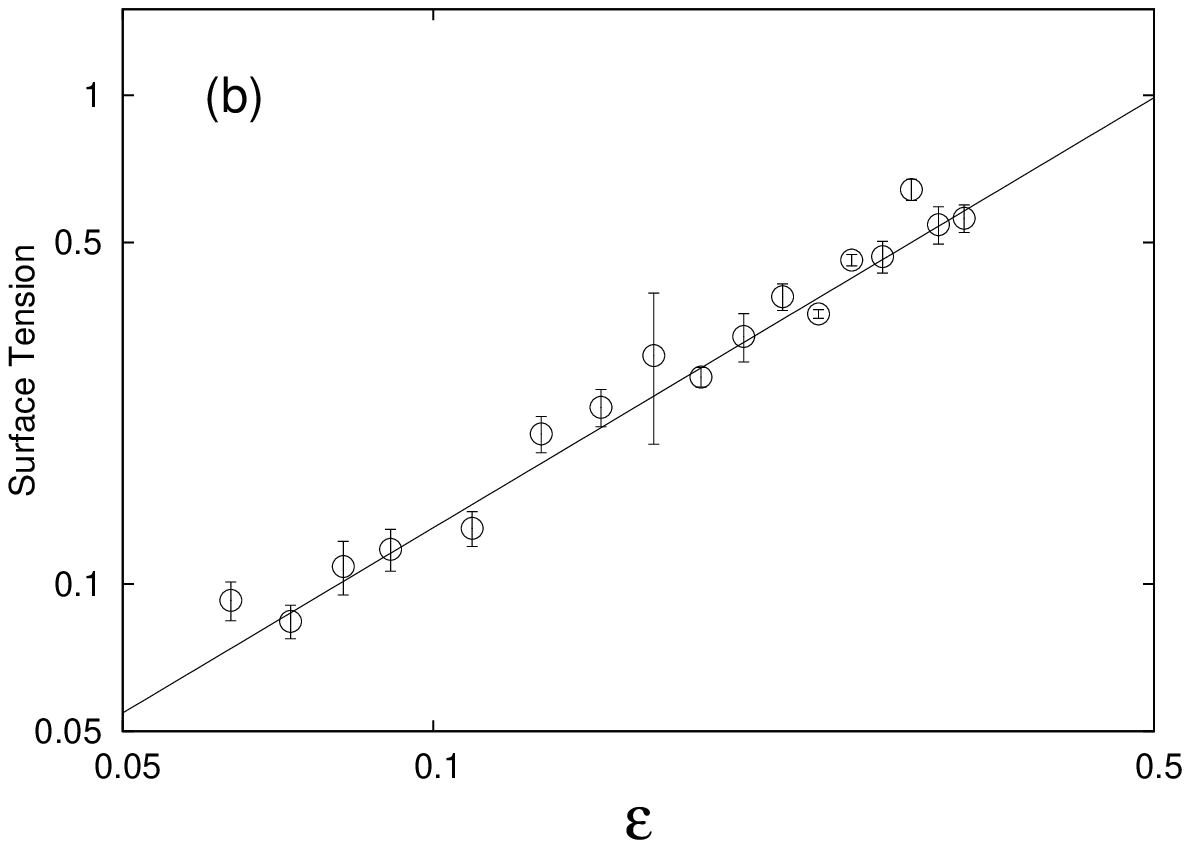}
\end{center}
\caption{(Color online) (a) Temperature dependence of the surface
thickness. The significant finite-size effect is observed.
(Inset) The interface broadening for $T=0.9$. Using Eq.~(\ref{eq_CWT}),
we can estimate the surface tension.
(b) Temperature dependence of the surface tension obtained from the
interface broadening behavior. The critical exponent is determined to be $\nu =0.63(4)$.
The solid line is the eye guide with the slope $2\nu = 1.26$.
}\label{fig_lambda}
\end{figure}

We also study the critical behavior of the interface thickness.
We find that the interface thickness exhibits strong finite-size
effect as shown in Fig.~\ref{fig_lambda}~(a),
while the finite-size effect on the the surface tension is almost negligible.
These finite-size dependencies come from the interface broadening
due to the capillary waves \cite{Werner1999, Vink2005}.
According to the capillary wave theory, the system size dependence
of the interface thickness $\lambda$ is expressed to be,
\begin{equation}
\lambda^2 = C + \frac{T}{4\gamma} \ln L, \label{eq_CWT}
\end{equation}
where $C$ is a system-size independent constant for a fixed temperature.
Note that, the temperature $T$ and system-size $L$ used here are dimensionless values.
The system-size dependence of the interface thickness for $T=0.9$
is shown in the inset of Fig.~\ref{fig_lambda}~(a).
From this finite-size dependence, we estimate the surface tension
as a function of temperature. The critical behavior of the surface tension
obtained from the interface thickness is shown in Fig.~\ref{fig_lambda}~(b).
The critical exponent of the correlation length is determined to be
$\nu = 0.63(4)$ which is consistent with the value of the Ising universality class.

\section{Discussions}

We determine the phase boundary between gas and liquid phases by
achieving gas-liquid coexisting state with MD. This method is
simple and provides the phase boundary between liquid and gas
phases with high accuracy. Since we have
gas-liquid interface in the system, we can check
the finite-size effect on the obtained results; if the linear
length of the system $L_z$ is much longer than
the interface thickness $\lambda$, then the coexisting densities of liquid and gas are
expected to be free from the finite-size effect. In our
experience, the finite-size effect on the coexisting density is
vanished when $L_z > 10 \lambda$. We determine the critical temperature
from the block density analysis of the Binder parameter, which allows us to
perform finite-size analysis only from one simulation run with
large system size. Additionally, this method does not require the
distribution of order parameter of Ising model at the criticality,
and therefore, we do not have to consider the mixed-field theory.

The critical exponents $\beta$ and $\nu$ are determined from the
behaviors of the order parameter and the surface tension. While
the critical exponents of the order parameter $\beta$ is
consistent with the Ising universality class, the exponent of
correlation length $\nu$ obtained from the surface tension
is found to be slightly larger than that of the  Ising universality.
While this deviation can be due to the capillary waves,
it is difficult to remove the effect since the surface tension does not
exhibit apparent finite-size effect.
Potoff and Panagiotopoulos studied the critical behavior of the surface
tension using HR technique and FSS analyses, and reported $\nu = 0.71(4)$ which is
much larger than the Ising universality~\cite{Potoff2000}.
However, they did not consider the effect from the capillary waves.
Considering the effect of the capillary waves, we obtained the
critical exponent of the correlation length $\nu = 0.63(4)$
which is consistent with the Ising universality. Note that, the statistical error became larger
than that of the value obtained directly from the stress tensor using Eq.~(\ref{eq_gamma}).
This is because the surface tension analysis with considering the capillary waves requires two steps;
first measuring the surface thicknesses for various system sizes and then extracting the
surface tension from its size dependence.
While the obtained values of exponents are consistent with that of the Ising universality class,
it is still difficult to declare that the universality class of the LJ system belongs to the Ising universality
since the accuracy is insufficient especially for the exponent of the correlation length.
Some techniques are required which can estimate the surface tension accurately.
Also the distances from the criticality is insufficient enough to discuss the universality class precisely.
In order to get closer to the critical point, larger and longer simulations are necessary.
It should be one of the further issues.

Some of the past studies reported the effect from the correction to
scaling~\cite{Wilding,Pellitero}. They first obtained the apparent
critical temperatures for several system size, and they took the
infinite size limit considering the correction. However, we did
not find any sign of the correction to scaling in
Fig.~\ref{fig_deltarho}. It is difficult to determine the value of
the correction to scaling exponent if it exists.
While the obtained values of critical exponents are consistent with the Ising universality
without considering the correction, it would become more important in order to
investigate more accurately.

Very recently, Ma and Hu
\cite{10MaHu1,10MaHu2,10MaHu3,10MaHu4,10HuMa} used MD to study
relaxation and aggregation of polymer chains, in which neighboring
monomers of a polymer chain are connected by rigid bonds
\cite{10MaHu1,10MaHu3} or springs with various spring constants
\cite{10MaHu2,10MaHu4}. They found that when the bending-angle
depending potential and the torsion-angle depending potential are
zero or very small, polymer chains tend to aggregate
\cite{10MaHu3,10MaHu4,10HuMa}. Such results are useful for
understanding mechanism of protein aggregation
\cite{10MaHu3,10MaHu4} related to neurodegenerative diseases,
including Alzheimer's disease (AD), Huntington's disease (HD),
Parkinson's disease (PD), etc. One may argue that polymer chains
can aggregate only when they are inside the phase boundary. It is
of interest to extend the method of the present paper to calculate
the phase diagram to calculate phase diagrams of polymer chains of
various chain lengths \cite{note}. Such phase diagrams will be
useful for understanding under what temperature and density
conditions, polymer chains would aggregate.

\section*{Acknowledgements}
The computation was carried out by using facilities of the
Supercomputer Center, Institute for Solid State Physics,
University of Tokyo, and Information Technology Center, Nagoya
University. We would like to thank K. Binder, N.
Kawashima, S. Todo, and Y. Tomita for helpful discussions. This
work was partially supported by Grants-in-Aid for Scientific
Research (Contract No.\ 23740287), by KAUST GRP (KUK-I1-005-04),
by Grants NSC 100-2112-M-001-003-MY2, and by NCTS (North).


\begin{thebibliography}{99}
\bibitem{71Gene} H. E. Stanley, {\it Introduction to Phase Transitions
 and Critical Phenomena} (Oxford Univ. Press, New York 1971).


\bibitem{90Leo}  L. P. Kadanoff,  Physica A, {\bf 163}, 1 (1990)


\bibitem{45jcp} E. A. Guggenheim, J. Chem. Phys. {\bf 13},
253(1945).


\bibitem{beta} More recent experimental results for liquid-gas
critical systems can be found in the paper V. Privman, P. C.
Hohenberg and A. Aharony in {\it Phase Transitions and Critical
Phenomena}: V. 14 edited by C. Domb and J.L. Lebowitz and C. Domb
(Academic Press, London, 1991).


\bibitem{09jcpBinder} B. M. Magnetti, {\it et al.},
J. Chem. Phys. {\bf 130}, 044101 (2009).
\bibitem{44ising} L. Onsager, Phys. Rev. {\bf 65}, 117(1944).
\bibitem{97ising} F. G. Wang and C.-K. Hu, Phys. Rev. E {\bf 56}, 2310 (1997).
\bibitem{sa94} D. Stauffer and A. Aharony, {\it Introduction to
   Percolation Theory}, Revised 2nd. ed. (Taylor and Francis, London, 1994).
\bibitem{95prl} C.-K. Hu, C.-Y. Lin, and J.-A. Chen, Phys. Rev. Lett.
   \textbf{75}, 193 (1995) and \textbf{75}, 2786(E) (1995); Physica A \textbf{
    221}, 80 (1995)
\bibitem{96prl} C.-K. Hu and C.-Y. Lin, Phys. Rev. Lett.
    \textbf{77}, 8 (1996).
\bibitem{04prl} H. Watanabe, S. Yukawa, N. Ito, and C.-K. Hu,
        Phys. Rev. Lett. \textbf{93}, 19601 (2004). This paper
        contains some typos, see Ref. \cite{05prlwh} for details;
        see also G. Pruessner and N. R. Moloney, Phys. Rev. Lett.
  {\bf 95}, 258901 (2005).
\bibitem{05prlwh} H. Watanabe and C.-K. Hu, Phys. Rev. Lett. \textbf{95}, 258902
(2005); Phys. Rev. E {\bf 78}, 041131 (2008).
\bibitem{dimer} N. Sh. Izmailian, V. B. Priezzhev, P. Ruelle, and C.-K. Hu,
Phys. Rev. Lett. {\bf 95}, 260602 (2005); N. Sh. Izmailian, K. B.
Oganesyan, M.-C. Wu, and C.-K. Hu, Phys. Rev. E {\bf 73}, 016128
(2006).
\bibitem{52pr-LeeYang} C. N. Yang and T. D. Lee, Phys. Rev. {\bf
87},  404 (1952); T. D. Lee and C. N. Yang, {\bf 87}, 410 (1952).
\bibitem{Watanabe2010} H. Watanabe, M. Suzuki, and N. Ito, Phys. Rev. E \textbf{82}, 051604 (2010).
\bibitem{Panagiotopoulos}  A. Z. Panagiotopoulos, Mol. Phys. \textbf{61}, 813 (1987);
A. Z. Panagiotopoulos, J. Phys.: Condens. Matter \textbf{12}, R25 (2000) .
\bibitem{Moeller} D. M\"oller and J. Fischer, Mol. Phys. \textbf{69}, 463 (1990).
\bibitem{Widom} B. Widom, J. Chem. Phys., \textbf{39}, 2808 (1963).
\bibitem{Okumura} H. Okumura and F. Yonezawa, J. Chem. Phys. \textbf{113}, 9162 (2000);
H. Okumura and F. Yonezawa, J. Phys. Soc. Jpn. \textbf{70}, 1990 (2001).
\bibitem{Wilding} N. B. Wilding and A. D. Bruce, J. Phys.: Condens. matter \textbf{4} 3087 (1992);
N. B. Wilding, Phys. Rev. E \textbf{52}, 602 (1995).
\bibitem{Pellitero} J. P\'erez-Pellitero, P. Ungerer, G. Orkoulas, and A. D. Mackie, J. Chem. Phys. 125, 054515 (2006).
\bibitem{96jpaIsing} A. L. Talapov and H. W. J. Bl$\ddot{o}$te, J. Phys. A: Math. Gen.
{\bf 29}, 5727  (1996).
\bibitem{Ito2000} N. Ito, K. Hukushima, K. Ogawa, and Y. Ozeki, J. Phys. Soc. Jpn. \textbf{69}, 1931 (2000).
\bibitem{Hasenbusch} M. Hasenbusch, Phys. Rev. B \textbf{82} 174433 (2010).
\bibitem{Spotswood1973} S. D. Stoddard and J. Ford, Phys. Rev. A \textbf{8}, 1504 (1973).
\bibitem{NoseHoover} W. G. Hoover,  Phys. Rev. A \textbf{31}, 1695 (1985).
\bibitem{Tuckerman1992} M. Tuckerman, B. J. Berne, and G. J. Martyna, J. Chem. Phys. \textbf{97}, 1990 (1992).
\bibitem{mdacp} http://mdacp.sourceforge.net/
\bibitem{mdnote} H. Watanabe, M. Suzuki, and N. Ito, Prog. Theor. Phys. \textbf{126}, 203-235 (2011).
\bibitem{McLennan} J. A. McLennan, \textit{Introduction to Non-equilibrium Statistical Mechanics} (Prentice Hall, 1996) .
\bibitem{Binder} K. Binder, Z. Phys. B: Condens. Matter \textbf{43}, 119 (1981).
\bibitem{Rovere} M. Rovere, D. W. Hermann, and K. Binder, Europhys. Lett., \textbf{6} 585 (1988).
\bibitem{rectilinear}
J. A. Zollweg and G. W. Mulholland, J. Chem. Phys. \textbf{57}, 1021 (1972);
A. B. Cornfeld and H. Y. Carr, Phys. Rev. Lett. \textbf{29}, 28 (1972);
A. Z. Panagiotopoulos, Int. J. Thermophys. \textbf{15}, 1057 (1994).
\bibitem{Wang2007} J. Wang and M. A. Anisimov, Phys. Rev. E \textbf{75}, 051107 (2007).
\bibitem{Pestak1987} M. W. Pestak, R. E. Goldstein, M. H. W. Chan, J. R. de Bruyn, D. A. Balzarini, and N. W. Ashcroft,
Phys. Rev. B \textbf{36} 599 (1987).
\bibitem{Werner1999} W. Werner, F. Schmid, M. M\"uller, and K. Binder, Phys. Rev. E, \textbf{59}, 728 (1999).
\bibitem{Vink2005} R. L. C. Vink, J. Horbach, and K. Binder, J. Chem. Phys., \textbf{122}, 134905 (2005).
\bibitem{Potoff2000} J. J. Potoff and A. Z. Panagiotopoulos, J. Chem. Phys., \textbf{112}, 6411 (2000).
\bibitem{10MaHu1} W.-J. Ma and C.-K. Hu, J. Phys. Soc. Jpn {\bf 79}, 024005 (2010).
\bibitem{10MaHu2} W.-J. Ma and C.-K. Hu, J. Phys. Soc. Jpn {\bf 79},  024006 (2010).
\bibitem{10MaHu3} W.-J. Ma and C.-K. Hu, J. Phys. Soc. Jpn {\bf 79},  054001  (2010).
\bibitem{10MaHu4} W.-J. Ma and C.-K. Hu, J. Phys. Soc. Jpn {\bf 79}, 104002 (2010).
\bibitem{10HuMa} C.-K. Hu and W.-J. Ma, Prog. Theor. Phys. Supp. {\bf 184}, 369 (2010).
\bibitem{note} Please note that Ref. \cite{09jcpBinder} contains
some phase diagrams of chain molecules.
\end{thebibliography}
\end{document}